\documentclass{article}
\usepackage{spconf,amsmath,epsfig}

\usepackage[]{graphicx}
\graphicspath{{Figures/}}
\usepackage{caption}
\usepackage{subfig}
\usepackage{amsmath}
\usepackage{amssymb}
\usepackage{epsfig}
\usepackage{cite}
\usepackage{color}
\usepackage{balance}
\usepackage{amsmath}
\usepackage{amsfonts} 
\usepackage{graphicx}
\usepackage{pdfpages}
\usepackage[T1]{fontenc} 
\usepackage{array}
\usepackage{url}

\usepackage{color}
\usepackage{wrapfig}
\usepackage{lipsum}
\usepackage[hidelinks]{hyperref}
\usepackage{multirow}
\usepackage{tabularx}
\usepackage{gensymb} 
\usepackage{breqn}
\usepackage{array}
\usepackage{booktabs}
\usepackage{bm}
\usepackage[linesnumbered,ruled,vlined]{algorithm2e}
\SetKwComment{Comment}{$\triangleright$\ }{}
\let\oldnl\nl
\newcommand{\nonl}{\renewcommand{\nl}{\let\nl\oldnl}}
\newcommand{\vars}{\texttt}

\title{An Extremely-low Cost Ground-Based Whole Sky Imager}

\name{Mayank Jain$^{1,2}$, Isabella Gollini$^{3}$, Michela Bertolotto$^{2}$, Gavin McArdle$^{2}$ and Soumyabrata Dev$^{1,2}$
\thanks{The research in this paper is funded by a research grant from Kaggle, a subsidiary of Google LLC. The ADAPT Centre for Digital Content Technology is funded under the SFI Research Centres Programme (Grant 13/RC/2106) and is co-funded under the European Regional Development Fund.}
\thanks{Send correspondence to S.\ Dev: \url{soumyabrata.dev@ucd.ie}}
}
\address{
	$^{1}$~The ADAPT SFI Research Centre, Ireland \\
	$^{2}$~School of Computer Science, University College Dublin, Ireland\\
	$^{3}$~School of Mathematics and Statistics, University College Dublin, Ireland
}

\begin{document}

\maketitle
\begin{abstract}
    Ground-based Whole Sky Imagers (WSIs) are increasingly being used for various remote sensing applications. While the fundamental requirements of a WSI are to make it climate-proof with an ability to capture high resolution images, cost also plays a significant role for wider scale adoption. This paper proposes an extremely low-cost alternative to the existing WSIs. In the designed model, high resolution images are captured with auto adjusting shutter speeds based on the surrounding light intensity. Furthermore, a manual data backup option using a portable memory drive is implemented for remote locations with no internet access.
\end{abstract}

\section{Introduction}
\label{sec:intro}
Images captured from whole sky imagers (WSIs) have been widely analyzed for various remote sensing applications such as radio-communications, climate analysis, and weather forecasting. These imagers provide images with high temporal and spatial resolution at low-cost as compared to their satellite counterparts~\cite{Dev2016GRSM, wang2018ground}. This makes them especially useful in the fields where short-term analysis is required, such as solar irradiance forecasting~\cite{solar_irr_pred, IGARSS_solar}.

While WSIs can generate a large amount of data, the generated images are often prone to high noise from sun glare, bird flocks, and dust particles. Thus, post processing of the generated images generally becomes an important pre-requisite. In such a scenario, a reduction in the apparatus design will not only helps to reduce the overall information retrieval cost but will also create a possibility to install multiple imagers for same locality in order to further reduce the noise. 

\begin{table}[!ht]
\centering
\begin{tabular}{ || m{6cm} | c || } 
\hline\hline
\textbf{Items} & \textbf{Cost (in \$)} \\
\hline\hline
Raspberry Pi 4 Model-B with 4GB RAM & 70 \\ 
\hline
Raspberry Pi High Quality Camera (with interchangeable lens base) & 70 \\
\hline
6mm Wide Angle Lens for Raspberry Pi High Quality Camera & 35 \\
\hline
Portable 1.5TB External Hard Drive & 60 \\
\hline
64GB microSDXC Memory Card & 10 \\
\hline
USB Type-C 15.3W Power Supply & 10 \\
\hline
DS3231 RTC Module & 3 \\
\hline
Duracell 2032 Coin Battery 3V & 2 \\
\hline
Micro HDMI to HDMI Cable (6 feet) & 7 \\
\hline
Metal Armour Case for Raspberry Pi 4 & 18 \\
\hline
Electronic Items: LDR, Push Button, LEDs, Resistances, and Breadboard & 4 \\
\hline
Polystyrene Insulating Ice Box & 3 \\
\hline
Plastic bottle and wooden pedestal for camera & 1 \\
\hline
Glass Dome & 1 \\
\hline
PVC Plyboard (1 sq. feet) & 1 \\
\hline
Electrical Accessories \& Other Items & 4 \\ 
\hline\hline
\textbf{Total Cost} & \textbf{299} \\ 
\hline\hline
\end{tabular}
\caption{Component description and cost analysis for designing the low-cost WSI}\vspace{-0.4cm}
\label{table:costTable}
\end{table}

An ideal WSI design must be climate proofed, widely implementable and should capture high resolution images. Different models have been developed for various industrial and research purposes. TSI~880~\cite{long2001total} and the panorama camera~\cite{chen2012howis} are being used for weather monitoring. UTSA~Sky Imager~\cite{richardson2017low}, WAHRSIS~\cite{WAHRSIS}, Canon~IXUS~II with FOV~180\degree ~\cite{kazantzidis2012cloud}, and Hemispheric Sky Imager~\cite{Long} are being used in cloud classification and wind data monitoring system applications. Monitoring of air traffic controls is another application of WSIs, an example of which is the IR~NEC~TS~9230 camera~\cite{infrared_UK}.

While TSI~880 is a commercialized model developed by Yankee Environmental Systems and costs around $\$30,000$, WSIs developed by individual research groups are generally cheaper like WAHRSIS with development cost of $\$1769$~\cite{WAHRSIS}. More recently, the UTSA~Sky Imager was released in $2017$ with strikingly low cost of $\$500$~\cite{richardson2017low}.

In this paper, we present a novel low-cost design of ground-based WSI based on the Raspberry Pi prototyping platform. Using much cheaper off-the-shelf alternatives, the design was built with an effective cost of under $\$300$. The list of detailed components and their prices is given in Table~\ref{table:costTable}.

\section{WSI Design and Construction}\label{sec:WSIdesign}
\subsection{Mechanical Design and Construction}\label{sec:mechDesign}

\begin{figure}[!ht]
    \centering
    \includegraphics[width=0.85\columnwidth]{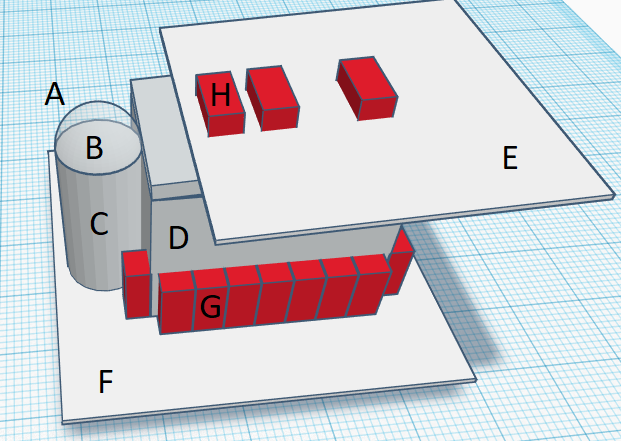}\\
    \includegraphics[width=0.85\columnwidth]{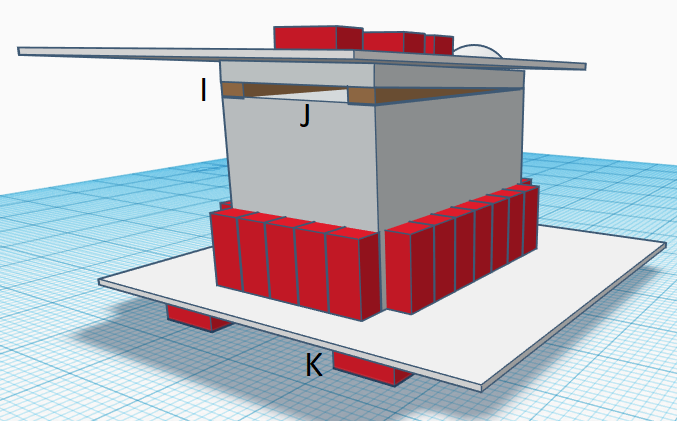}
    \caption{Mechanical design of the WSI. \textit{A}: glass dome; \textit{B}: camera assembly \& LDR; \textit{C}: camera holder and pedestal; \textit{D}: insulating thermocol ice box; \textit{E, F}: PVC plyboard; \textit{G, H}: bricks to keep apparatus in place; \textit{I}: small wedges; \textit{J}: raised pedestal with bricks}
    \label{fig:elecFrameworkWSI}
\end{figure}

A polystyrene-based insulation box (which is generally used to store dry ice) was placed on a pedestal to prevent the basic electronic setup from outside heat and rain. The box contains Raspberry Pi, hard drive, breadboard circuit, battery and switch. LEDs were placed in the breadboard circuit to monitor basic performance of hard drive and WSI without attaching a screen in the setup. The Raspberry Pi was enclosed in a metal armored case that comes with two fans and thermal tapes to prevent the Pi from heating. To ensure efficient outflow of heat generated by the components inside the polystyrene box, small wedges (shown by I label in Figure 1) were left on one side. Further, to protect our electronic setup from rain, a large PVC Plyboard was placed on top of the box with it's edges lying way outside the boundary of the box.

The camera was mounted on a wooden pedestal inside an empty plastic bottle. A hole was made in the bottle cap to make space for lens of the camera. Glass dome was fixed with cement, paint, and fast setting epoxy resin and hardener on the hollow cap. It was done to make the camera dome water resistant and protect the camera from rain and other contaminants. To adjust proper shutter settings, a light dependent resistance (LDR) was placed alongside the camera in the dome.

To connect the camera and LDR (inside the plastic bottle unit) with the electronic section (in the polystyrene box unit), slits were made in the plastic bottle and polystyrene box. Then, the two units were taped together to prevent separation and breakage. To further prevent rain and moisture from penetrating through the slits, a polyethylene based thin protective cover was used. It was fixed with the bottle cap and the lid of polystyrene box in order to cover the slits in the middle. Finally, the whole setup was protected by bricks from all sides to strengthen it.

\begin{figure}[!ht]
    \centering
    \includegraphics[width=0.75\columnwidth]{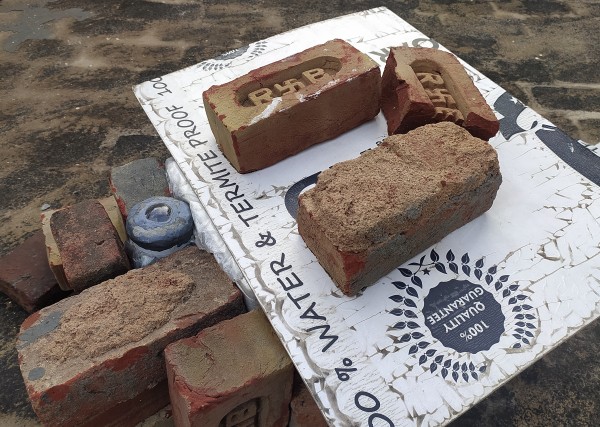}\vspace{0.1cm}\\
    \includegraphics[angle=270,origin=c,width=0.75\columnwidth]{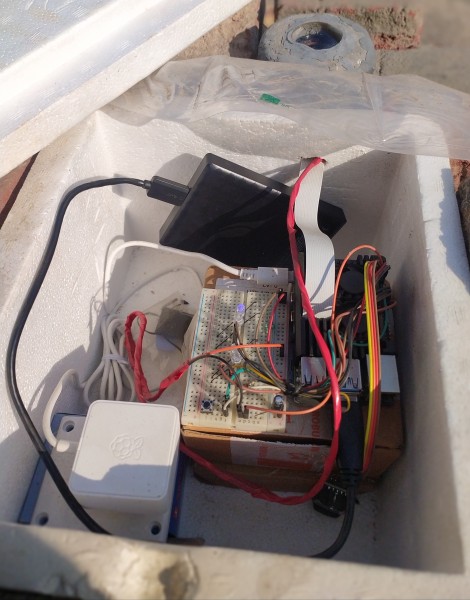}\vspace{-0.8cm}
    \caption{Functional form of the designed low cost WSI}
    \label{fig:functionalWSI}
    \vspace{-0.5cm}
\end{figure}

\subsection{Electronic Framework \& Logic Handling}\label{sec:elecFrameworkWSI}

The electronic framework for the WSI is designed over the Raspberry Pi prototyping platform (as shown in Fig.~\ref{fig:elecFrameworkWSI}). At the software end, we have bifurcated the system into -
\begin{itemize}
    \setlength\itemsep{0em}
    \item the camera handling part, and
    \item the data handling part
\end{itemize}

\begin{figure}[!ht]
    \centering
    \includegraphics[width=0.9\columnwidth]{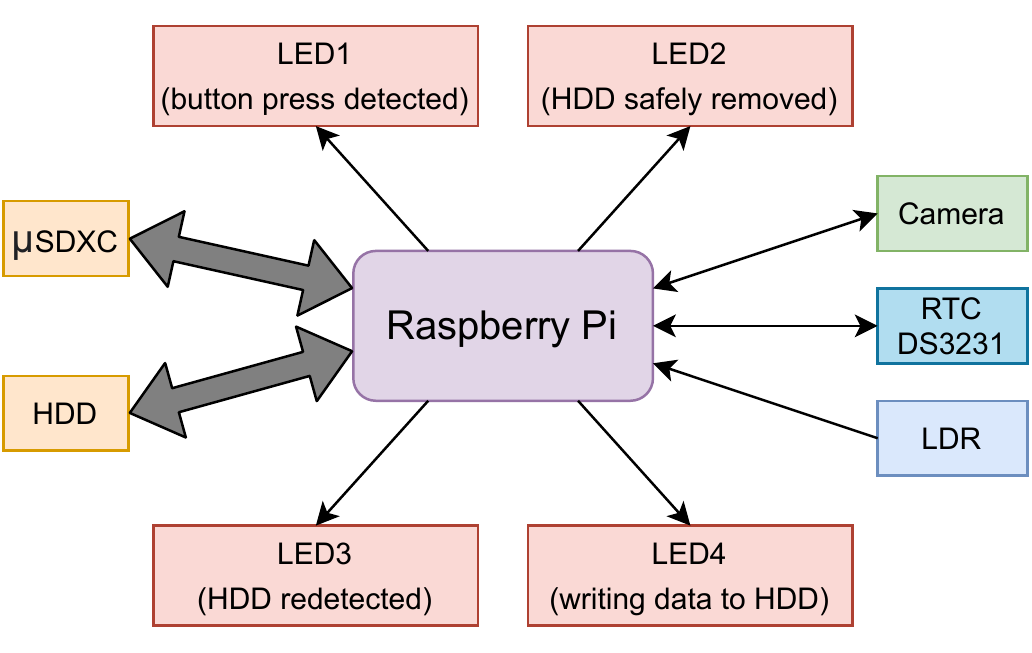}
    \caption{Electronic framework of the designed low-cost WSI}
    \label{fig:elecFrameworkWSI}
    \vspace{-0.1cm}
\end{figure}

The camera handling part is responsible for capturing the images periodically. It is managed via a cron job. While the main script is tasked only to capture the image and save it in the local memory ($\mu$SDXC), the cron job is responsible to execute the script after every $5$ minutes. Since the cron job is managed by the Linux operating system itself, it ensures the fail-safe operation of the camera. The main script looks at the values of LDR and adjusts the shutter time of the camera accordingly. If the LDR value is too low, it means night and hence the shutter is allowed to remain open for a longer time than usual in order to get a better view of the night sky.

Since the local memory ($\mu$SDXC) is very small ($64$GB), the camera needs to store images elsewhere in order to be able to run for a longer period of time. To keep the camera completely portable and easy to install at remote locations, we provided the system with a $1.5$TB (enough to store images for more than $600$ days) external memory (HDD). At $1200$ hours and $2400$ hours every day, the system checks for the presence of the external HDD and if it is available, it transfers all the data from the $\mu$SDXC to the HDD. The HDD can be manually removed and reinstalled to take a second backup of the captured images. To facilitate safe removal, $4$ LEDs and a push button is provided alongside the system. The complete process of HDD management is defined in the Algorithm~\ref{algo:HDDmanagement}. Since the main code for HDD management script runs in infinite loop, it is configured as daemon and the script's running status is checked and maintained by another cron job.

\setlength{\textfloatsep}{15pt}
\begin{algorithm}[!ht]
$\vars{pb} \gets \text{handle for push button}$\\
$\vars{l}_{\vars{1}}, \vars{l}_{\vars{2}}, \vars{l}_{\vars{3}}, \vars{l}_{\vars{4}} \gets \text{handles for 4 LEDs}$\\
$\vars{HDDstate} \gets TRUE$\\
$\vars{ct()} \gets$ utility function to fetch current system time\\
$\vars{hs()} \gets$ utility function to fetch exact HDD state\\
\Comment{\textbf{Interrupt Service Routine:}}
\If{$state(\vars{pb}) == pressed$}{
    $state(\vars{l}_{\vars{1}}) = ON$\;
    safely eject HDD\;
    $state(\vars{HDDstate}) = FALSE$\;
    $state(\vars{l}_{\vars{2}}) = ON$\;
    $wait(5s)$\;
    $state(\vars{l}_{\vars{1}}, \vars{l}_{\vars{2}}) = OFF$\;
    }
\Comment{\textbf{Main Code (run in infinite loop):}}
\eIf{$\vars{HDDstate}$ and \vars{ct()} == (1200 or 2400)}{
    $state(\vars{l}_{\vars{4}}) = ON$\;
    transfer data from $\mu$SDXC to HDD\;
    $state(\vars{l}_{\vars{4}}) = OFF$\;
    }{
    \If{not $\vars{HDDstate}$ and $\vars{hs()}$}{
        $\vars{HDDstate} = TRUE$\;
        $state(\vars{l}_{\vars{3}}) = ON$\;
        $wait(5s)$\;
        $state(\vars{l}_{\vars{3}}) = OFF$\;
        }
    }
\caption{HDD management daemon}
\label{algo:HDDmanagement}
\end{algorithm}

\section{Results \& Limitations}\label{sec:results}
\begin{figure*}
    \centering
    \includegraphics[width=0.23\textwidth]{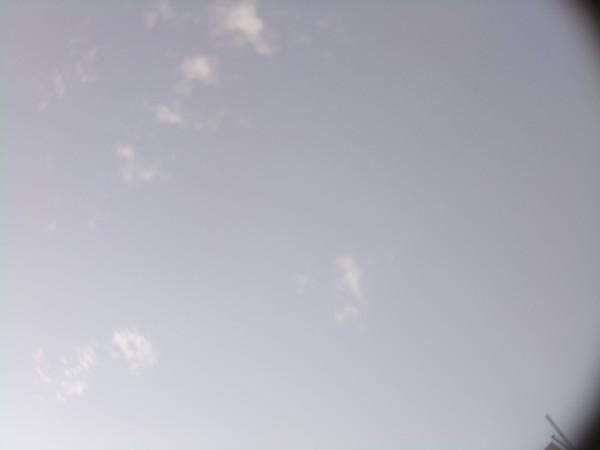}
    \includegraphics[width=0.23\textwidth]{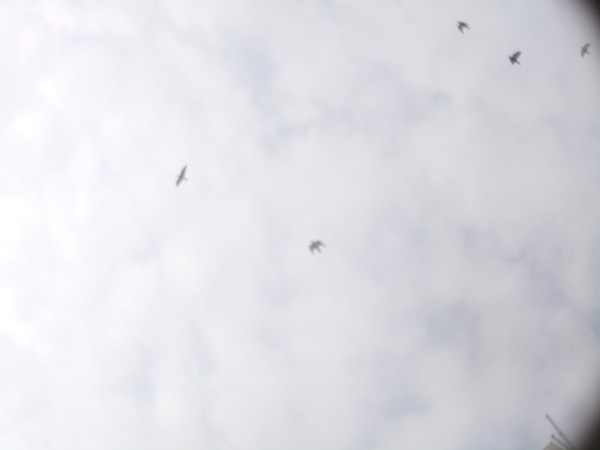}
    \includegraphics[width=0.23\textwidth]{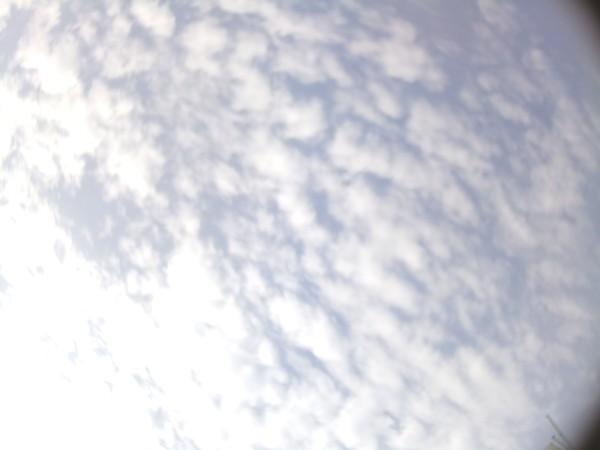}
    \includegraphics[width=0.23\textwidth]{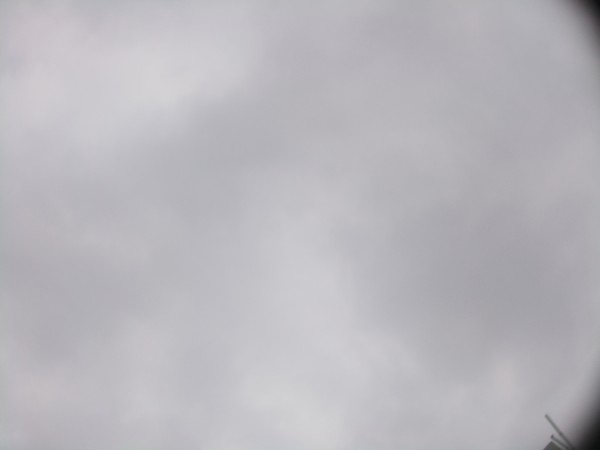}\\
    \includegraphics[width=0.23\textwidth]{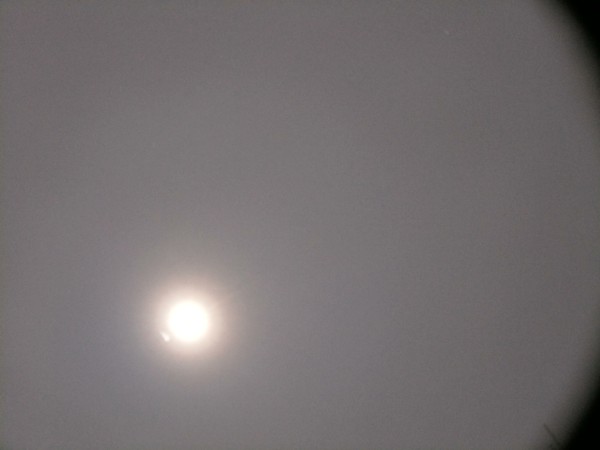}
    \includegraphics[width=0.23\textwidth]{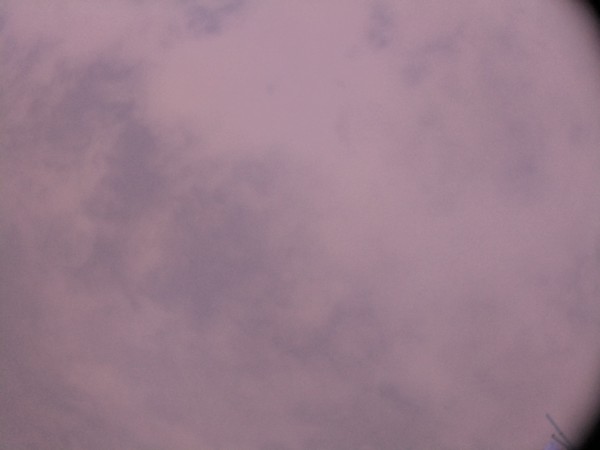}
    \includegraphics[width=0.23\textwidth]{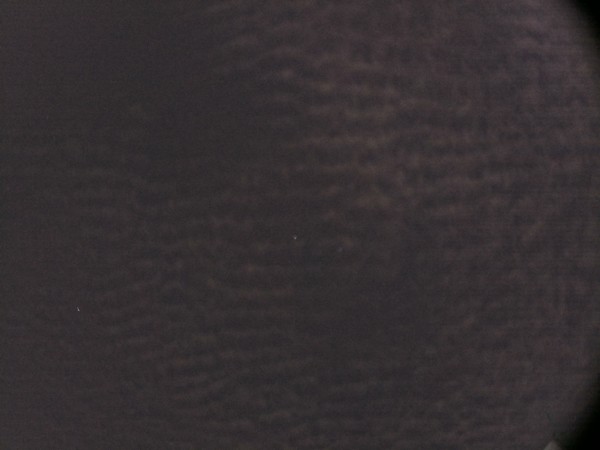}
    \includegraphics[width=0.23\textwidth]{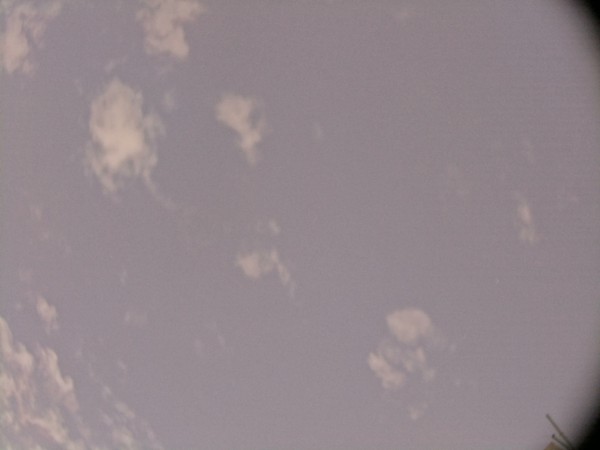}
    \caption{Sample images that were captured by the designed WSI. \textit{Starting from top left corner in a clockwise manner -} clear sky during day; thick white clouds during day; patterned clouds during day; thick dark clouds during day; almost clear sky at dusk; patterned clouds during night; thick white clouds during night; and clear sky during night (full-moon).}
    \label{fig:WSIcaptuedImages}
\end{figure*}

The designed WSI is capturing sky/cloud images for over $6$ months now with only a few instances of failure. Figure~\ref{fig:WSIcaptuedImages} shows some of the captured images during different times of the day. Salient features of the designed model can be summarized as follows:
\vspace{-0.5em}
\begin{itemize}
    \setlength\itemsep{-0.4em}
    \item Captures images in high resolution while automatically adjusting the shutter speed enabling it to capture images in low-light conditions
    \item Local backup facility with easy handling of hard drive to avoid dependence on internet
    \item Deploys RTC to keep track of time during power outage
    \item Durable and low-cost chassis which is well-protected from outside heat and other weather conditions
\end{itemize}

While the cloud cover can be identified clearly in the captured images, some of them yet suffer from the following problems:
\begin{itemize}
    \setlength\itemsep{-0.4em}
    \item Unrealistic coloration of the images due to infra-red rays that are being captured by the camera
    \item Sun glare coming in through the corner of images making it hard to identify the cloud cover accurately
    \item Dust particles and water vapors sometimes get deposited over the glass dome (which is protecting the camera) resulting in blurry images
    \item Although in a rare case, the camera sensor gets heated up and fails to capture any image whatsoever
\end{itemize}
\vspace{-0.2cm}

\section{Conclusions \& Future Work}
\label{sec:conc}
This paper presents an extremely low-cost design for a ground-based WSI using readily available components. The camera is capable of capturing high-resolution images ($4056\times3040$) with varying shutter speeds based on the LDR readings. With the capability to capture images at all times of the day and at a regular interval of $5$ minutes, the WSI is suitable for various applications including (but not limited to) correlation studies with meteorological variables to assist in forecasting of solar irradiance and rainfall events. Furthermore, the additional feature of creating local backup makes the device ultra portable and thereby suitable for remote locations where internet is generally not accessible. 

In future, we plan to reduce the limitations of the current design. To resolve the issue of rising temperature of the camera sensor, we are planning to add a cooling device for the camera as we have for the Raspberry Pi within the box. Further, adding an IR-filter underneath the lens of the camera is planned for automatic color correction in the images. Lastly, a defogger/heating element might be added to the glass dome so that the condensed water droplets and/or ice can be removed automatically from it in order to obtain clear images. Although, the modifications will increase the overall cost of the WSI, the increment is not expected to be a significant one.

\balance


\end{document}